\documentclass[a4paper]{jpconf}
\usepackage[utf8]{inputenc}
\usepackage{graphicx}
\usepackage{amsmath}
\usepackage{amssymb}

\begin{document}
\title{Application of the random matrix theory to the boson peak in glasses}

\author{D A Conyuh$^{1,2}$, Y M Beltukov$^1$, D A Parshin$^3$}
\address{$^1$Ioffe Institute, 194021 Saint-Petersburg, Russia}
\address{$^2$Peter the Great St.Petersburg Polytechnic University, 195251 Saint-Petersburg, Russia}
\address{$^3$St.Petersburg Academic University, 199034 Saint-Petersburg, Russia}

\ead{conyuh.dmitrij@yandex.ru}

\begin{abstract}
    The density of vibrational states $g(\omega)$ of an amorphous system is studied by using the random-matrix theory. Taking into account the most important correlations between elements of the random matrix of the system, equations for the density of vibrational states $g(\omega)$ are obtained. The analysis of these equations shows that in the low-frequency region the vibrational density of states has the Debye behavior $g(\omega) \sim \omega^2$. In the higher frequency region, there is the boson peak as an additional contribution to the density of states. The obtained equations are in a good agreement with the numerical results and allow us to find an exact shape of the boson peak.
\end{abstract}

\section{Introduction}
The finding of common vibrational properties in amorphous dielectrics (glasses) is one of the key problems in the physics of disordered systems. One of these properties is a boson peak in the reduced density of vibrational states~\cite{Gurevich 2003}. It characterizes a presence of the excess vibrational density of state compared to the Debye law.

The boson peak is observed in many experiments, such as the Raman scattering, X-ray scattering, an inelastic neutron scattering, and the measurements of the heat capacity $C(T)$ as a maximum in $C(T)/T^3$. The boson peak frequency $\omega_b$ is correlated with Ioffe-Regel crossover frequency $\omega_\textsc{ir}$ when the phonon mean free path becomes of the order of the phonon wavelength~\cite{Gurevich-1993, Parshin-2001, Ruffle-2006, Ruffle-2008, Shintani-2008}. Despite of a number of articles about the boson peak, still there is no common physical interpretation. In the present paper, to study the nature of the boson peak we apply an approach based on the use of random matrices.

\section{Random matrix model}
In amorphous solids, due to the local disorder, the symmetric force constant matrix $\hat{\Phi}$ has a random nature. Therefore, the dynamical matrix $M_{ij}=(m_im_j)^{-1/2}\Phi_{ij}$~\cite{Maradudin 1971} has some disorder as well. However, the dynamical matrix $\hat{M}$ has several physical constraints. First of all, there is a translational invariance (or a sum rule), which means that the sum of all elements in each row or column of the matrix $\hat{M}$ must be equal to zero (we consider unit masses $m_i=1$). Secondly, eigenvalues of the dynamical matrix $\hat{M}$ determine the eigenfrequencies squared and must be nonnegative~\cite{Beltukov 2011} due to the requirement of the mechanical stability.

Taking into account the above properties, the dynamical matrix $\hat{M}$ can be written as $\hat{M} = \hat{A}\hat{A}^T$~\cite{Beltukov 2013}. The matrix $\hat{A}$ is some non-symmetric random matrix with a constraint $\sum_i A_{ij}=0$. In general, the matrix $\hat{A}$ can be a rectangular $N \times K$ random matrix.

Let us consider a disordered system on a simple cubic lattice with the lattice constant $a_0=1$. In this model each atom is placed in a site of the lattice, but the interaction between atoms is random. Such a system can be built by the following structure of the matrix $\hat{A}$. The non-diagonal elements $A_{ij}$ are Gaussian random numbers if atoms with indices $i$ and $j$ are nearest neighbors in the lattice. Other non-diagonal elements are zeros. The diagonal elements are defined as $A_{ii} = -\sum_{j\neq i}A_{ji}$ to satisfy the condition of the translational invariance.

As it was shown in~\cite{Beltukov 2013, Beltukov 2016}, the system, described by the square matrix $\hat{A}$, has a zero Young modulus. The point is that each column of $\hat{A}$ corresponds to one bond of the system, each row of $\hat{A}$ corresponds to one degree of freedom of the system. Therefore, for the square matrix $\hat{A}$, the number of bonds and the number of degrees of freedom are equal. This is the so-called Maxwell rule. In this case, the system is very soft and do not have a macroscopic rigidity. In such a system, phonons do not propagate, and the boson peak is absent too~\cite{Beltukov 2013, Beltukov 2016}.

In order to introduce the macroscopic rigidity and the boson peak into the matrix model $\hat{M}=\hat{A}\hat{A}^T$, we build a new random rectangular matrix $\hat{B}$ with size $N \times K$ with $K > N$. We take two independent realizations of the square random matrix $\hat{A}$ with size $N \times N$: $\hat{A}$ and $\hat{A'}$. In order to generate the matrix $\hat{B}$, we randomly insert some number of columns of the matrix $\hat{A'}$ into the matrix $\hat{A}$. This random insertion of the new columns corresponds to a random addition of new bonds to the vibrational system. The relative number of new bonds is $\kappa = (K - N)/N$, where $K$ is the total number of columns in the resulting matrix $\hat{B}$. Therefore, the new rectangular matrix $\hat{B}$ satisfies conditions of translational invariance and mechanical stability and describes the vibrational system as $\hat{M} = \hat{B}\hat{B}^T$.

The analysis of the boson peak with a help of the random matrix theory can shed light on its universal nature. However, the nontrivial structure of the matrix $\hat{B}$ significantly complicates the problem. Nevertheless, the model under consideration has simpler correlations between matrix elements than three similar random matrix models, which were considered in~\cite{Beltukov 2013, Beltukov 2016}. In the next section, we present equations for finding the vibrational density of states (VDOS) $g(\omega)$, obtained with an account of the most important correlations between elements of the matrix $\hat{B}$.

\section{Density of vibrational states}
The pairwise correlations between matrix elements in the matrix $\hat{B}$ has a form
\begin{equation}
    \langle B_{ij}B_{kl} \rangle = \frac{1}{N}C_{ik}\delta_{jl},  \label{eq:Bcor}
\end{equation}
where the averaging is performed by different realizations of the matrix $\hat{B}$ with the fixed parameter $\kappa$. The matrix $\hat{C}$ has the following structure. The non-diagonal elements $C_{ij}=-2$ if atoms with indices $i$ and $j$ are nearest neighbors in the lattice. Otherwise, $C_{ij} = 0$. Diagonal elements are $C_{ii}=12$. One can see that the matrix $\hat{C}$ can be considered as a dynamical matrix of a perfect crystal.

The distribution of eigenvalues of the correlation matrix $\hat{C}$ can be written as
\begin{multline}
    \rho(\varepsilon) = \frac{1}{8\pi^3}\int_{-\pi}^{\pi}\int_{-\pi}^{\pi}\int_{-\pi}^{\pi} \delta\left(\varepsilon - 8\sin^2\frac{k_x}{2} - 8\sin^2\frac{k_y}{2} - 8\sin^2\frac{k_z}{2}\right) dk_x dk_y dk_z\\
    = \int_0^\infty J_0^3(\xi) \cos(3\xi\varepsilon) d\xi, \quad \label{eq:rho}
\end{multline}
where $J_0(\xi)$ is the Bessel function of zero order.

\begin{figure}[h!]
    \centering
    \includegraphics[scale=0.7]{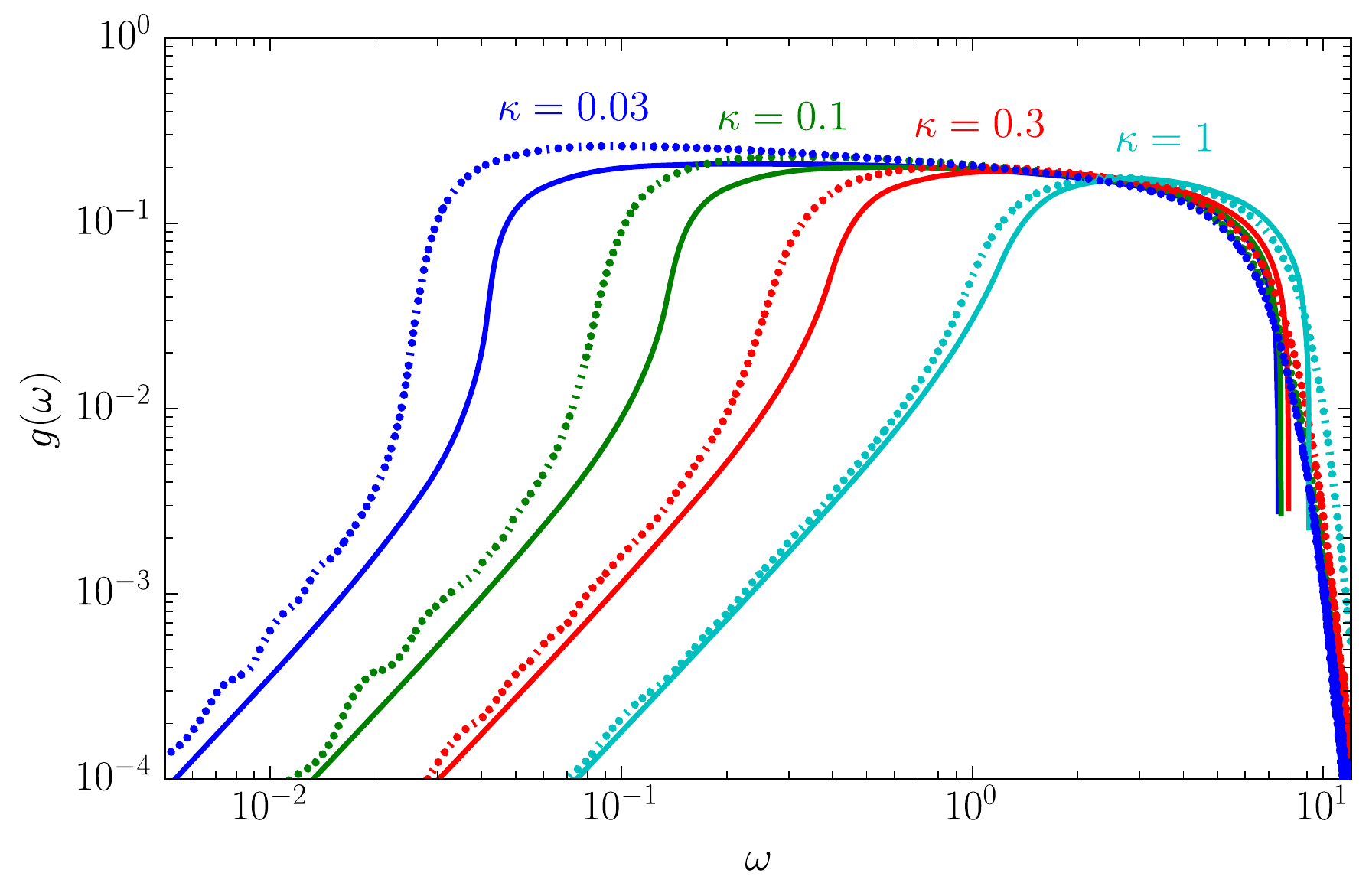}
    \caption{The vibrational density of states in the ensemble $\hat{M} = \hat{B}\hat{B}^T$ for different parameters $\kappa$. Solid lines show the solution of Eqs.~(\ref{eq:rho})--(\ref{eq:VDOS}). Dotted lines were computed by KPM for a numerical realization of the matrix $\hat{B}$ in a system with $N=400^3$ atoms.}
    \label{fig:Density}
\end{figure}

If we take into account the pairwise correlations only (Eq.~(\ref{eq:Bcor})), the VDOS can be obtained from the following system of equations~\cite{Burda 2004, Burda 2006}:
\begin{gather}
    \int \frac{\rho(\varepsilon) \varepsilon^2}{(X-\varepsilon)^2 + Y^2} d\varepsilon = 1+\kappa, \label{eq:crit_hor} \\
    \omega^2=(X^2+Y^2) \int \frac{\rho(\varepsilon) \varepsilon}{(X-\varepsilon)^2 + Y^2} d\varepsilon,
    \label{eq:frec} \\
    g(\omega) = \frac{1}{\pi}\frac{2\omega Y}{X^2 + Y^2}. \label{eq:VDOS}
\end{gather}
Equation~(\ref{eq:crit_hor}) defines a so-called critical horizon, that is a contour in the complex plane $Z = X+iY$.  Equations (\ref{eq:frec}) and (\ref{eq:VDOS}) yield an implicit dependency between $g(\omega)$ and $\omega$ from the critical horizon. Figure \ref{fig:Density} shows
the
VDOS of the ensemble $\hat{M} = \hat{B}\hat{B}^T$. Solid lines show a numerical solution of Eqs.~(\ref{eq:rho})--(\ref{eq:VDOS}). One can see a crossover between the low-frequency Debye law $g(\omega) \sim \omega^2$ and approximately constant VDOS at high frequencies. The crossover frequency $\omega_c$ depends on the value of the parameter~$\kappa$.

In Fig.~\ref{fig:Density} dotted lines show the result of the Kernel Polynomial Method (KPM)~\cite{Beltukov 2016 PRE,Weisse 2006} applied to
a
randomly generated matrix $\hat{B}$ for a system with $N=400^3$ atoms. It has a qualitative agreement with   the solution of Eqs.~(\ref{eq:rho})--(\ref{eq:VDOS}) for the same values of $\kappa$. The quantitative difference can be caused by neglecting high-order correlations between elements of the matrix $\hat{B}$.

\section{Low-frequency asymptotics}

Let us consider the low-frequency asymptotics of the VDOS $g(\omega)$ for $\kappa \ll 1$. One can show that this region is defined by the region $X\ll 1$ and $Y \ll 1$ of the critical horizon (\ref{eq:crit_hor}). In this case we can assume that $\rho(\varepsilon) = \sqrt{\varepsilon/128\pi^4} + {\cal O}(\varepsilon^{3/2})$ for $\varepsilon\ll 1$. In this case, we can write Eqs.~(\ref{eq:crit_hor}) and (\ref{eq:frec}) in the following form:
\begin{gather}
    2XI_1 + \frac{(X^2-Y^2)}{16\pi Y}\sqrt{X+\sqrt{X^2+Y^2}} = \kappa, \label{eq:crit_hor_simple}\\
    \omega^2=(X^2+Y^2)\left(I_1 + \frac{X}{16\pi Y}\sqrt{X+\sqrt{X^2+Y^2}} \right), \label{eq:frec_simple}
\end{gather}
where
\begin{equation}
    I_1 = \int d\varepsilon \frac{\rho(\varepsilon)}{\varepsilon} = \frac{1}{4}\int_0^\infty J_0(\zeta)\sin(3\zeta)d\zeta  \approx 0.126365.
\end{equation}
Below we consider two cases: $Y \ll X \ll 1 $ and $X \sim Y \ll 1$. As we will see, the first case is the left part of the boson peak, including low-frequency Debye behavior. The second case is the right part of the boson peak, which includes the maximum of $g(\omega)/\omega^2$.

\subsection{Left part of the boson peak}

In the case $Y \ll X \ll 1$, we neglect $Y^2$ compared to $X^2$ in Eqs.~(\ref{eq:crit_hor_simple}), (\ref{eq:frec_simple}) and (\ref{eq:VDOS}) and obtain the following system of equations:
\begin{gather}
    2XI_1 + \frac{X^2}{16\pi Y}\sqrt{2X} = \kappa, \label{eq:crit_hor_c1}\\
    \omega^2 = X^2\left(I_1 + \frac{X}{16\pi Y}\sqrt{2X} \right), \label{eq:frec_c1}\\
    g_l(\omega) = \frac{1}{\pi}\frac{2\omega Y}{X^2}. \label{eq:VDOS_c1}
\end{gather}
From the above equations, we find the VDOS which corresponds to the left part of the boson peak:
\begin{equation}
    g_l(\omega) = \frac{\omega}{8\pi^2\sqrt{I_1\kappa}} \sqrt{\frac{1-\sqrt{1-\omega^2/\omega_c^2}}{1-\omega^2/\omega_c^2}}, \label{eq:VDOS_lp}
\end{equation}
where $\omega_c = \kappa/(2\sqrt{I_1})$ is the crossover frequency. One can show that $|g_l(\omega) - g(\omega)| \ll g(\omega)$ in the region $\omega < \omega_c - \delta$, where $\delta\sim \kappa^{3/2}$.

In the low-frequencies region $\omega \ll \omega_c$, the VDOS (\ref{eq:VDOS_lp}) has the Debye behavior:
\begin{equation}
     g_D(\omega) = \frac{\omega^2}{2\pi^2(2\kappa)^{3/2}} + {\cal O}(\omega^{4}). \label{eq:VDOS_Debye}
\end{equation}

\subsection{Right part of the boson peak}

\begin{figure}[h!]
    \centering
    \includegraphics[scale=0.64]{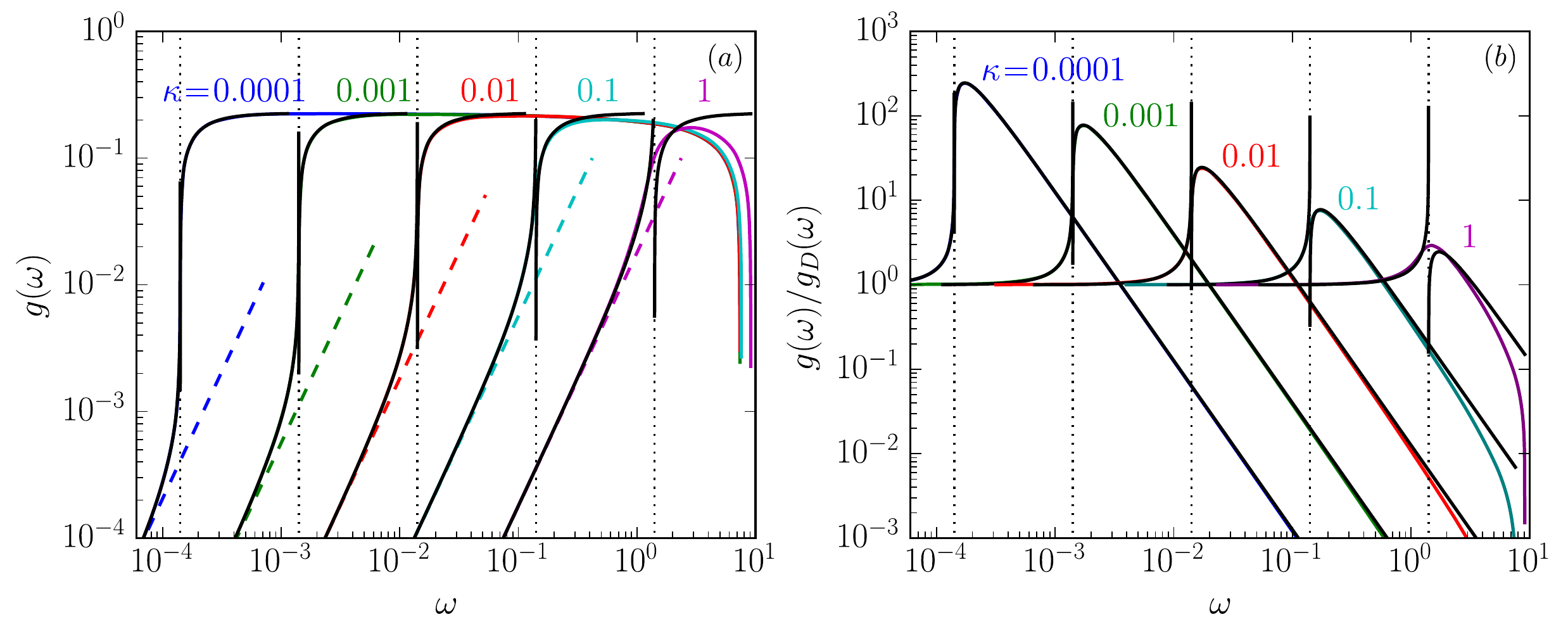}
    \caption{(a) The vibrational density of states in the ensemble $\hat{M} = \hat{B}\hat{B}^T$ for different parameters $\kappa$. Color solid lines show the solution of Eqs.~(\ref{eq:rho})--(\ref{eq:VDOS}). Color dashed lines show the Debye VDOS $g_D(\omega)$. Black solid lines show asymptotic expressions $g_l(\omega)$ and $g_r(\omega)$. Vertical dotted lines show the crossover frequency $\omega_c$. (b) The same for the reduced VDOS $g(\omega)/g_D(\omega)$.}
    \label{fig:VDOS}
\end{figure}

In the case $X \sim Y \ll 1$, the critical horizon (\ref{eq:crit_hor_simple}) is approximately vertical line $X \approx X_c = \kappa/(2\sqrt{I_1})$. Therefore, Eqs. (\ref{eq:frec_simple}) and (\ref{eq:VDOS}) can be written as
\begin{gather}
    \omega^2 = (X_c^2+Y^2)I_1, \label{eq:frec_c2}\\
    g_r(\omega) = \frac{1}{\pi}\frac{2\omega Y}{X_c^2 + Y^2}. \label{eq:VDOS_c2}
\end{gather}
From the above  equations, we find the VDOS, which corresponds to the right part of the boson peak:
\begin{equation}
    g_r(\omega) = \frac{2\sqrt{I_1}}{\pi}\sqrt{1 - \frac{\omega_c^2}{\omega^2}}. \label{eq:VDOS_rp}
\end{equation}
One can show that $|g_r(\omega) - g(\omega)| \ll g(\omega)$ in the region $\omega > \omega_c + \delta$, where $\delta\sim \kappa^{3/2}$.  As a result, the asymptotics $g_l(\omega)$ and $g_r(\omega)$ almost coincide with $g(\omega)$ in the full low-frequency region $\omega \ll 1$ except a narrow crossover region $\omega_c - \delta < \omega < \omega_c + \delta$ (Fig.~\ref{fig:VDOS}a).

As can be seen in Fig.\ref{fig:VDOS}b, there exists a boson peak in the reduced VDOS $g(\omega)/g_D(\omega)$. For small parameters $\kappa$, two asymptotic expressions (\ref{eq:VDOS_lp}) and (\ref{eq:VDOS_rp}) have a good agreement with the precise solution of Eqs.~(\ref{eq:rho})--(\ref{eq:VDOS}). In the narrow region $\omega_c - \delta < \omega < \omega_c + \delta$ there is a crossover between $g_l(\omega)$ and $g_r(\omega)$.

\section{Young modulus and boson peak frequency}
It was found that the boson-peak frequency $\omega_b$ (frequency, where $g(\omega)/g_D(\omega)$ has a maximum) depends on the Young modulus $E$ of the system as $\omega_b \sim E$~\cite{Wyart 2010,Kojima 1999,Beltukov 2016}. We can find the relation between $\omega_b$ and $E$ from the obtained asymptotics (\ref{eq:VDOS_lp}) and (\ref{eq:VDOS_rp}).

In the simple cubic lattice with unit lattice constant $a_0=1$ and unit masses $m_i=1$, the Debye's law has a form
\begin{equation}
    g_D = \frac{1}{2\pi^2} E^{-3/2} \omega^2. \label{eq:Debye_law}
\end{equation}
Comparing Eqs.~(\ref{eq:VDOS_Debye}) and (\ref{eq:Debye_law}), we find that
\begin{equation}
     E = 2\kappa = 2\left(\frac{K}{N}-1\right). \label{eq:E}
\end{equation}
As it follows from this form, if the matrix $\hat{B}$ is square, Young modulus $E$ is equal to zero. Indeed, according to the Maxwell rule, a system has no macroscopical rigidity when the number of bonds is equal to the number of degrees of freedom. The result for the Young modulus (\ref{eq:E}) is valid only for small $\kappa$, i.e. sides of matrix $\hat{B}$ are not much different. In other words, the system should be soft enough, but have a non-zero stiffness, which can be calculated from (\ref{eq:E}) in the random matrix model.

Figure \ref{fig:VDOS} shows that the boson-peak frequency $\omega_b$ lies on the right asymptotic $g_r(\omega)$. In order to find expressions for $\omega_b$, we find the derivative of $g_r(\omega)/g_D(\omega)$ with respect to $\omega$ and equate it to zero. As a result, we obtain the boson-peak frequency
\begin{equation}
    \omega_b = \sqrt{\frac{3}{2}}\omega_c =  \sqrt{\frac{3}{8I_1}}\kappa. \label{eq:BP}
\end{equation}
From Eqs.~(\ref{eq:E}) and (\ref{eq:BP}) we find the linear relation between the boson peak on the Young modulus:
\begin{equation}
    \omega_b = \sqrt{\frac{3}{32I_1}}E. \label{eq:EBP}
\end{equation}

\section{Conclusion}

Summarizing, in the present paper we derived closed analytical expressions to find the VDOS $g(\omega)$ of the random matrix model of an amorphous system. The solution of these equations for $g(\omega)$ agrees with the result of the direct calculation of the VDOS by the KPM. Also, from these equations, we found the crossover, which determines the shape of the boson peak. The boson peak frequency $\omega_b$ is close to the crossover frequency $\omega_c = \sqrt{2/3}\,\omega_b$. The VDOS to the left and to the right of $\omega_c$ are described by the asymptotic expressions, $g_l(\omega)$ and $g_r(\omega)$, respectively. The low-frequency behavior of $g_l(\omega)$ exhibits the Debye law $g(\omega) \sim \omega^2$ and gives the Young modulus $E \sim \kappa \sim \omega_b$. The linear relation between the Young modulus and the boson peak has been found in different systems~\cite{Wyart 2010,Kojima 1999,Beltukov 2016} but was not explained so far.

{\ack The authors are thankful to V. I. Kozub for his constructive suggestions and comments. This work is supported by the Russian Federation President Grant no. MK-3052.2019.2.}

\end{document}